\documentclass{article}

\def\xxinput#1{\input#1}

\xxinput{vsolj02.sty}

\usepackage{longtable}
\usepackage[dvipdfmx]{graphicx}
\usepackage[comma,colon]{natbib}

\usepackage[OT2,T1]{fontenc}

\def\cite{\citealt}
\setcitestyle{aysep={}}

\newcounter{author}
\def\altaffilmark#1{$^{#1}$}
\def\altaffiltext#1{$^{#1}$\,}

\def\authorcount#1#2{{\refstepcounter{author}\label{#1}
                     \altaffiltext{\ref{#1}}{#2}}}

\begin{document}

\begin{center}

\title{Very low state in PY Per in 2022}

\author{
        Taichi~Kato\altaffilmark{\ref{affil:Kyoto}}
}
\email{tkato@kusastro.kyoto-u.ac.jp}

\authorcount{affil:Kyoto}{
     Department of Astronomy, Kyoto University, Sakyo-ku,
     Kyoto 606-8502, Japan}

\end{center}

\begin{abstract}
\xxinput{abst.inc}
\end{abstract}

   In \citet{kat22pyper}, I reported the detection of
an outburst (2021 December--2022 January) resembling
an SU~UMa-type superoutburst in the Z Cam star PY Per
with an orbital period of
0.15468(5)~d \citep{kat22pyper,tay96arandamcaspyper}.
\citet{kat22pyper} also detected faint states in
observations by the American Association of
Variable Stars (AAVSO), and showed that this object
is also an VY Scl-type cataclysmic variable (CV)
[see e.g., \citet{war95book} for CVs in general and
their subtypes].
VY Scl-type faint states are sometimes associated with
Z Cam stars, most notably the 1996--1997 totally
unexpected long-lasting fading in
RX And \citep{sio01rxand,kat02rxand,sch02rxand,kat04rxandsuuma}.

\begin{figure*}
\begin{center}
\includegraphics[width=16cm]{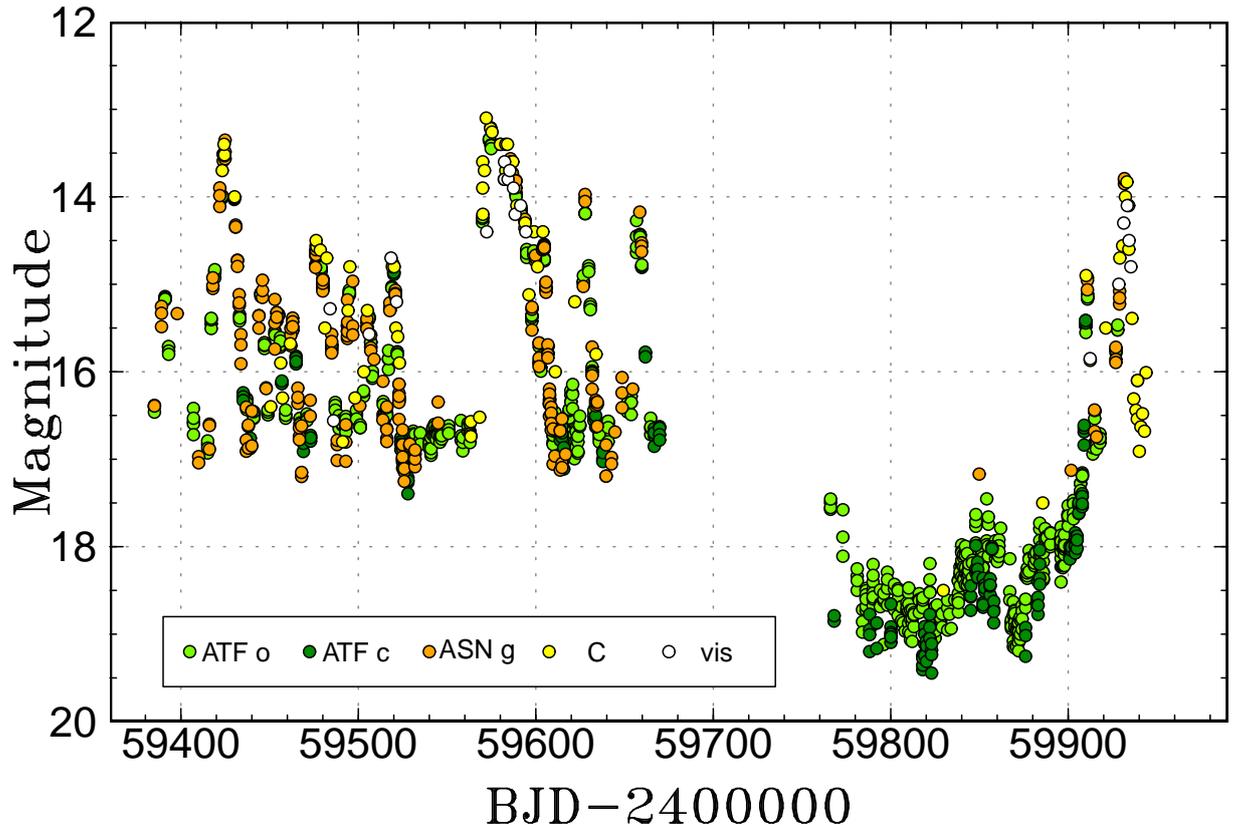}
\caption{
  2021--2022 light curve of PY Per using
  ATLAS forced photometry (ATF), ASAS-SN (ASN),
  VSOLJ and VSNET
  (C for CCD close to visual and vis for visual) observations.
  The left part of this figure corresponds to the fourth
  panel of Fig. 1 in \citet{kat22pyper}.  The outburst
  resembling an SU UMa-type superoutburst started on
  BJD 2459570.  After the solar conjunction, the object
  was found already in a deep, low state below 18~mag.
  The object gradually started to brighten after BJD 2459880
  and there was a short outburst on BJD 2459910.
  A more ordinary outburst started on BJD 2459927, which
  slowly rose to a maximum of 13.8~mag.
}
\label{fig:pyperlc}
\end{center}
\end{figure*}

   After the solar conjunction following
the unusual outburst mentioned above, I noticed that
PY Per was in low state without any outburst
in Variable Star Observers League
in Japan (VSOLJ) and VSNET \citep{VSNET} observations
(vsnet-alert 27029.\footnote{
  $<$http://ooruri.kusastro.kyoto-u.ac.jp/mailarchive/vsnet-alert/27029$>$
}).
This has been confirmed by using the All-Sky Automated Survey
for Supernovae (ASAS-SN)
Sky Patrol \citep{ASASSN,koc17ASASSNLC} data.
These observations started in 2022 June (ASAS-SN)
and 2022 July (VSOLJ).
Using the data by the Asteroid Terrestrial-impact
Last Alert System (ATLAS: \cite{ATLAS,hei18ATLASvar,smi20ALTASserver})
Forced Photometry
\citep{shi21ALTASforced}\footnote{
  The ATLAS Forced Photometry is available at
  $<$https://fallingstar-data.com/forcedphot/$>$.
}, I further confirmed that the 2022 fading episode was
the deepest (reaching 19.1 mag) and longest
(more than 150~d) ever recorded in this object
(figure \ref{fig:pyperlc}).

\begin{figure*}
\begin{center}
\includegraphics[width=16cm]{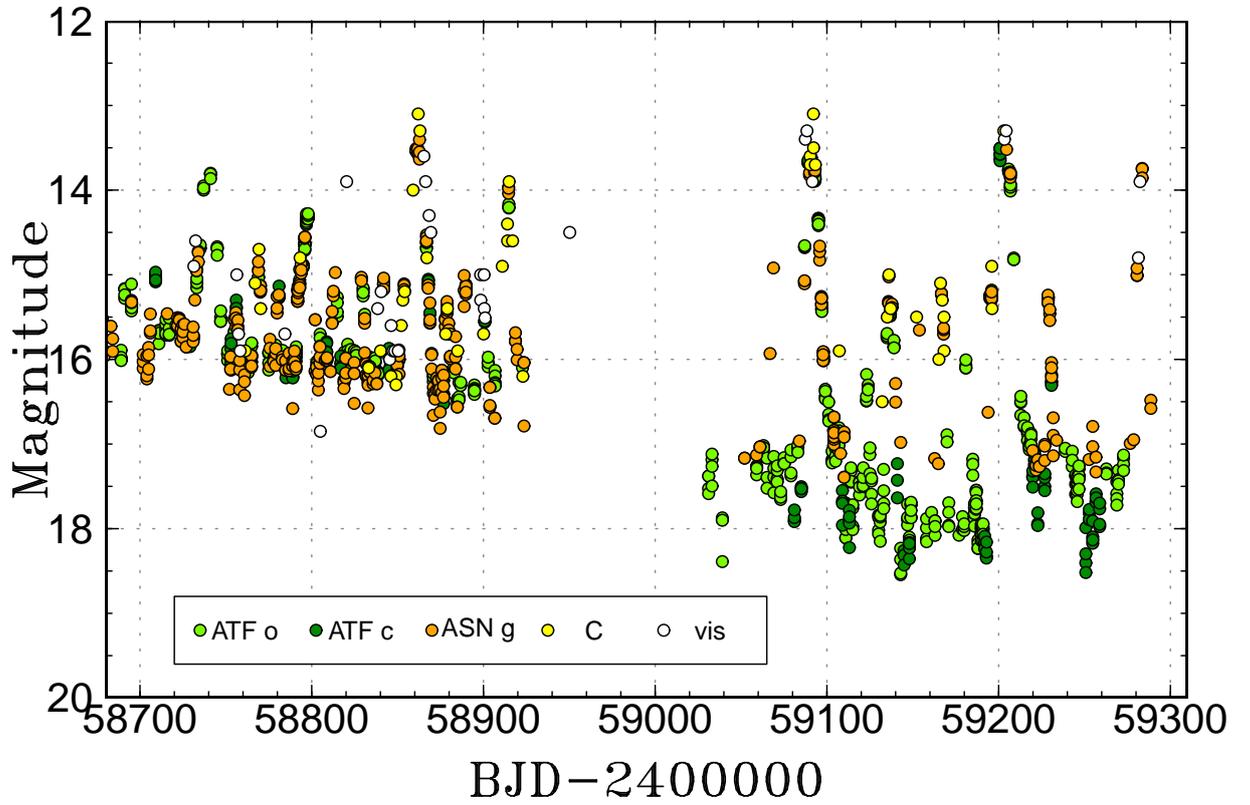}
\caption{
  Light curve of PY Per in the 2019--2021 season.
  The symbols are the same as in figure \ref{fig:pyperlc}.
  In the 2019--2020 season (before BJD 2458950), the object
  showed frequent low-amplitude outbursts as was typical for
  a Z Cam star.
  In the 2020--2021 season (BJD 2459030--2459290),
  the quiescence became fainter (near 18.0~mag) and
  longer and brighter outbursts in addition to smaller ones
  became more prominent than in the previous season.
}
\label{fig:pyperlc2}
\end{center}
\end{figure*}

   For a comparison, a light curve of the preceding seasons
is given in figure \ref{fig:pyperlc2}.
In the 2019--2020 season (before BJD 2458950), the object
showed frequent low-amplitude outbursts as was typical for
a Z Cam star.  The behavior was similar to this at least
between 2016 and 2019.
In the 2020--2021 season (BJD 2459030--2459290),
the quiescence became fainter (near 18.0~mag) and
longer and brighter outbursts in addition to smaller ones
became more prominent than in the previous season.
Although the quiescent brightness (18.0~mag) was almost
as faint as the VY Scl-type low state mentioned
in \citet{kat22pyper}, the object definitely showed
dwarf nova-type outbursts.  The light curve in 2020--2021
probably illustrated the behavior when the mass-transfer
rate decreased.

   Compared to the 2020--2021 season, the mass-transfer
rate probably returned normal in the 2021--2022 season
(left part of figure \ref{fig:pyperlc}).
There was, however, an interval lacking outbursts
(BJD 2459525--2459568), but not as faint as the 2020--2021
quiescence, preceding the unusual outburst resembling
an SU UMa-type superoutburst.  Accumulation of matter
in the disk during this interval may have caused
the unusual outburst.  Although this unusual outburst
may have been physically related to the very faint low state
in 2022, observations were impossible due to the solar
conjunction and how this very faint state started remains
a mystery.  Note that the 1996--1997 low state in RX And
was preceded by an unusually long standstill.  These rare
phenomena might have been physically related and the case
would also be suspected in PY Per.

\begin{figure*}
  \begin{center}
    \includegraphics[width=16cm]{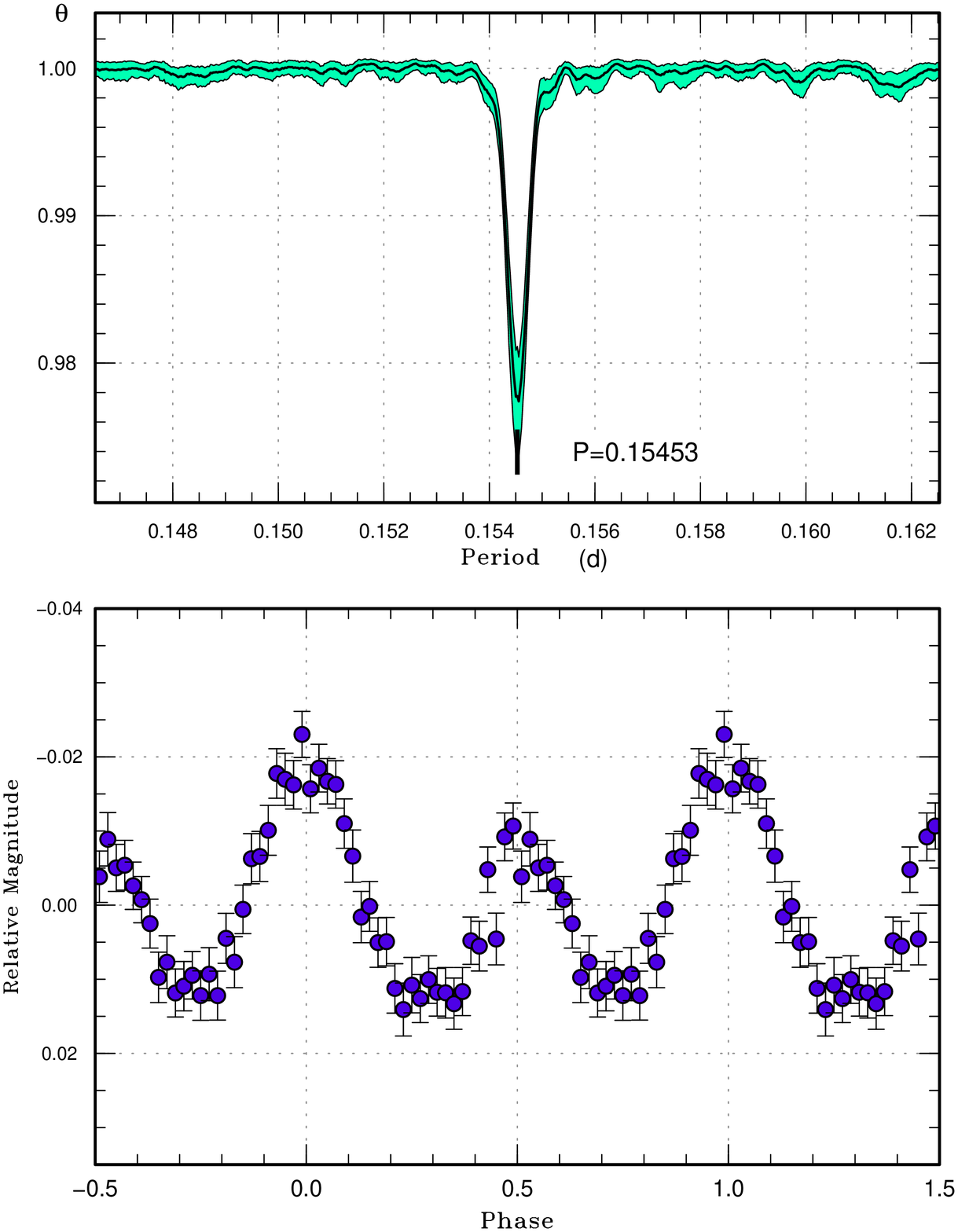}
  \end{center}
  \caption{Period analysis of the TESS data during the low state.
  (Upper): We analyzed 100 samples which randomly contain 50\% of
  observations, and performed the PDM analysis for these samples.
  The bootstrap result is shown as a form of 90\% confidence intervals
  in the resultant PDM $\theta$ statistics.
  (Lower): Orbital variation.  Two peaks in one orbit are clearly
  visible.  Note that the amplitude was smaller than real due to
  the contaminating stars.
  }
  \label{fig:pdm}
\end{figure*}

   I analyzed Transiting Exoplanet Survey Satellite (TESS)
observations obtained in 2022.\footnote{
  $<$https://tess.mit.edu/observations/$>$.
}  The full light-curve
is available at the Mikulski Archive for Space Telescope
(MAST\footnote{
  $<$http://archive.stsci.edu/$>$.
}).  The TESS observations started on 2022 October 28
(BJD 2459882) and ended on 2022 November 26 (BJD 2459910).
PY Per started rising in the TESS data on the final two days.
Since most of the TESS data were obtained when PY Per
was in deep low state and since the object has nearby
(unrelated) contaminating stars, I did not attempt to
extract the flux of PY Per but used the flux combined with
contaminating stars.  Using the data before BJD 2459908.6
(object in low state), I could detect the orbital period
and modulations (figure \ref{fig:pdm}).  The period was
determined to be 0.15453(2)~d with the Phase Dispersion Minimization
(PDM, \cite{PDM}) method after removing long-term trends
by locally-weighted polynomial regression
(LOWESS: \cite{LOWESS}).
The errors of periods by the PDM method were
estimated by the methods of \citet{fer89error} and \citet{Pdot2}.
Although the obtained period is in agreement with that obtained
in high state in \citet{kat22pyper}, the orbital profile
is very different.  In high state, there was a single peak
in one orbit [figure 3 in \citet{kat22pyper}],
while the current observations clearly show two maxima
in one orbit.  This feature most likely represents
an ellipsoidal variation of the secondary and TESS photometry
supports the very weak (or no) contribution from
the accretion disk.  These observations support
that the mass accretion almost stopped in this very low state
in PY Per and strengthen the identification of this object
as a VY Scl star.

\section*{Acknowledgements}

This work was supported by JSPS KAKENHI Grant Number 21K03616.
The author is grateful to the ASAS-SN, ATLAS and TESS teams
for making their data available to the public.
I am grateful to VSOLJ and VSNET observers for
reporting observations and to Naoto Kojiguchi for
providing a script for downloading TESS data.
The contributors from VSNET and VSOLJ in 2022
were Pavol A. Dubovsky, Masao Funada, Hiroshi Itoh,
Eddy Muyllaert, Yutaka Maeda, Masayuki Moriyama and Gary Poyner.

This work has made use of data from the Asteroid Terrestrial-impact
Last Alert System (ATLAS) project. The Asteroid Terrestrial-impact
Last Alert System (ATLAS) project is primarily funded to search for
near earth asteroids through NASA grants NN12AR55G, 80NSSC18K0284,
and 80NSSC18K1575; byproducts of the NEO search include images and
catalogs from the survey area. This work was partially funded by
Kepler/K2 grant J1944/80NSSC19K0112 and HST GO-15889, and STFC
grants ST/T000198/1 and ST/S006109/1. The ATLAS science products
have been made possible through the contributions of the University
of Hawaii Institute for Astronomy, the Queen's University Belfast, 
the Space Telescope Science Institute, the South African Astronomical
Observatory, and The Millennium Institute of Astrophysics (MAS), Chile.

\section*{List of objects in this paper}
\xxinput{objlist.inc}

I provide two forms of the references section (for ADS
and as published) so that the references can be easily
incorporated into ADS.

\renewcommand\refname{\textbf{References (for ADS)}}

\newcommand{\noop}[1]{}\newcommand{\hyphalt}{-}

\xxinput{pyperlowaph.bbl}

\renewcommand\refname{\textbf{References (as published)}}
\xxinput{pyperlow.bbl.vsolj}

\end{document}